\documentstyle[11pt,aaspp4,flushrt]{article}
\begin{document}

\title{The Halo from the Outside In: QSO Absorption 
Lines\altaffilmark{1}}
\pagestyle{empty}

\author{Christopher~W.~Churchill and Steven~S.~Vogt}
\affil{UCO/Lick Observatories, University of California, Santa Cruz}
\altaffiltext{1}{to appear in {\it Formation of the Galactic Halo...
Inside and Out}, eds.~H.~Morrison \& A.~Sarajedini 
(PASP Conference Series)}

\begin{abstract}
Studies of high resolution quasar absorption lines (QALs), arising 
from gas in local and $z \leq 1.5$ galaxies, provide
{\it direct}\/ probes of the kinematic, chemical, and ionization 
conditions and evolution of Milky Way--like galaxies.  Inferences 
drawn from these studies provide a powerful analog for those based
upon direct studies of the Halo.  The observed gas kinematics and 
velocity evolution reveal that galactic halos are complex dynamical 
components, consistent with Searle--Zinn formation scenarios.
\end{abstract}

\section*{Introduction}
\pagestyle{myheadings}
\markboth{\sc Churchill \& Vogt \hfill The Halo: QSO Absorption~~}
         {\sc Churchill \& Vogt \hfill The Halo: QSO Absorption~~}

As Sidney van den Bergh stated in his closing comments of the meeting,
there are two general approaches to studying the history and
evolution of the Galactic Halo.  
The first is to study the kinematic, chemical, and structural
components of the Halo itself, and the second is to study the 
halos of external galaxies with the partial aim of placing the Halo in
the broader context offered by various morphologies and environments
(cf.~Morrison, this volume).
Generally, the tracers of Halo formation (e.g.~globular clusters,
RR~Lyrae stars, Blue Horizontal Branch stars, star counts, etc.) are
treated as frozen relics of past formation processes.
The hope is that these processes have left signatures, such as
kinematic trends or metallicity gradients, which can be used to
clearly discern between competing formation scenarios.

One important tracer, which to a certain degree has not captured the
attention of those studying Milky Way evolution {\it per se},
is low density gas in the Halo.
Using the absorption lines seen in the spectra of background quasars
(QSO Absorption Lines, or QALs), several groups have carefully mapped
out the sky locations of large cloud complexes (cf.~van~Woerden, this
volume), including high--velocity clouds (HVCs), which by themselves
have a sky covering factor of 38\%.
The connection between HVCs and Halo tracers is not known, though
Majewski (this volume) has reported stellar moving groups which
correlate well with the locations and velocities of HVCs toward
the North Galactic Pole.
From QAL surveys of galaxies at redshifts $0.4 \leq z \leq
2.2$, gaseous halos of normal galaxies are known to extend to $\sim
70$~kpc and to contain an estimated $10^{9}$--$10^{10}$~M$_{\sun}$ of
gas (Steidel 1993; Steidel \& Sargent 1992).
Halo gas is the reservoir for star formation and chemical evolution, 
and plays a central role in the formation of the tracers used to study 
Halo formation. 
Its study in early epoch galaxies promises to yield important
clues to the processes that regulated galaxy formation
(cf.~Bechtold, this volume).

\section*{Learning about the Milky Way from QALs}

QAL studies are unique in that they directly probe galactic gas over
the entire history of galaxy evolution.
Thus, they provide a powerful method from which to indirectly
``view'' Milky Way Halo formation over a multi--billion year period
and offer a very broad context within which studies using Halo tracers
can be placed.
In tandem with high spatial resolution imaging (Hubble Space Telescope)
and high quality spectra of the absorbing galaxies themselves, the
potential provided by QAL studies can be fully realized.

From high resolution QAL spectra (see Fig.~1), we can measure the
number of clouds intercepted in a galaxy, their column densities,
broadening mechanisms, and line--of--sight velocities.   
HST images provide the absorbing galaxies' environments, luminosities,
colors, impact parameters to the QSO (projected galactocentric
distance of the absorbing clouds), and orientations relative to the
line of sight.
Spectra of the galaxies can be used to estimate star formation rates,
and (with LRIS on Keck) obtain rotation curves out to $z \sim 1$.
Case by case, we can study the physical details of absorbing gas and
its relationship to the host galaxy, and then piece together a
comprehensive picture of halo evolution directly from the large range of
epochs the galaxies sample.
The unexplored connections between absorption properties and the
locations probed in galaxies, their morphologies, redshifts, and
environments will ultimately be used to develop a global picture of
kinematic and chemical evolution.

As a first step, we have observed 24 QSOs with the HIRES spectrograph
(Vogt et~al.~1994). 
We have resolved absorption profiles of the Mg~{\sc ii}
$\lambda\lambda2796,2803$ resonant doublet in $\sim 50$ intervening
galaxies.
Many of these galaxies have been ground--based imaged in the
IR/optical and spectroscopically confirmed to have the redshift seen
in absorption (Steidel 1995; Steidel, Dickinson, \& Persson 1996).


\section*{Gaseous Fragments and Kinematic Evolution}

In this contribution, we present partial results and a brief discussion
of work to be published elsewhere (Churchill \& Vogt 1996).
In Fig.~1, we show four HIRES/Keck Mg~{\sc ii} absorption profiles
as seen in the spectra of background QSOs.
These absorption lines arise in low ionization gas, which also gives
rise to Mg~{\sc i} and Fe~{\sc ii} transitions.
Generally, these profiles appear to exhibit ``characteristics''
related to the location and structure probed by the QSO line of sight.
In particular, we note the complex high velocity spread in the
$z=0.51$ galaxy toward Q1254 and the $z=0.92$ galaxy toward Q1206.
One could interpret the optically--thick components of these profiles
as arising from the disks of these galaxies.
The HVC--like optically--thin components are more difficult to
understand.
However, they are highly suggestive of a picture in which material in
galaxy halos is comprised of kinematically and physically distinct
clumps, consistent with the Searle--Zinn (1978) picture of halo formation.


There is a  similarity between the features of the $z=1.17$ profile
toward Q1421 and that of the optically--thick components of Q1206.
These two QALs may arise in similar structures, or parts of
the galaxies.
The $z=1.55$ system toward Q1213 may be a merging or double galaxy,
the strength variations in the ``double'' profile being due to
the different line--of--sight orientations, morphologies, and/or
masses of the galaxies.
HST imaging would be decisive in testing these conjectures.
Such inferences can be drawn from QAL studies of local galaxies.
Bowen, Blades, \& Pettini (1995) have observed a ``double'' profile
similar to that of Q1213 that samples a line of sight passing
through both M81 and the Milky Way and spans 400~km~s$^{-1}$.
Churchill, Vogt, \& Steidel (1995) have observed a possible double
galaxy at $z=0.74$ that exhibits a richly structured ``double''
profile spanning 300~km~s$^{-1}$.
In a ground--based image, there are two galaxies of nearly equal
magnitude (redshift?), each with impact parameter $\sim 20$~kpc.

In Fig.~2, we present the probability, $P(\Delta v)$, of observing any
two clouds with line--of--sight velocity difference $\Delta v$.
The cloud--cloud velocity dispersion within halos exhibits strong
redshift evolution, becoming tighter as redshift decreases such that
by a look--back time of $\sim$ 8--10$h^{-1}$~Gyr
($h=H_0/100$~km~s$^{-1}$ Mpc$^{-1}$) Mg~{\sc ii} absorbing clouds have
a mean velocity dispersion $\sim 60$~km~s$^{-1}$.
Such {\it pronounced}\/ evolution may be due to biasing in
our sample selection, since we targeted an absorber population known
to exhibit evolution (Steidel \& Sargent 1992).
Since there is no apparent evolution in the sizes of halos (Steidel \&
Sargent 1992), or the number of clouds (this study), the implications
are that an observable shift may be occurring in the mechanisms giving
rise to significant amounts of high velocity material as early as
$\sim 8$ billion years ago.
Perhaps we are seeing the epoch at which the frequency of dwarf
satellite galaxy accretion onto their primary slows considerably. 

\acknowledgments
We would like to thank the Organizing Committee for hosting an
enjoyable meeting and a small travel grant for CWC.
Thanks to Jane Charlton, Ken Lanzetta, and Chuck Steidel for 
insightful on--going discussions.
This work has been supported in part by the Sigma Xi Grants--in--Aid of
Research program, the California Space Institute, and NASA
grant NAGW--3571.

\end{document}